\documentclass[superscriptaddress,twocolumn,10pt,pra,aps,reprint,showpacs,longbibliography]{revtex4-1}
\usepackage{float}
\floatstyle{boxed}
\usepackage{wrapfig}
\usepackage{array}
\usepackage{graphicx}
\usepackage{amsbsy}
\usepackage{xcolor}
\usepackage[utf8x]{inputenc}
\usepackage{epstopdf}
\usepackage{amsmath,amssymb,amsfonts}
\usepackage{indentfirst}
\usepackage{soul}
\usepackage[T1]{fontenc}

\newcommand{\PT}{${\cal PT}$~}

\renewcommand{\eqref}[1]{\mbox{Eq.~(\ref{#1})}}

\newcommand{\be}{\begin{equation}}
\newcommand{\ee}{\end{equation}}
\newcommand{\bea}{\begin{eqnarray}}
\newcommand{\eea}{\end{eqnarray}}
\newcommand{\Hef}{\hat H_{\rm eff}}

\usepackage{url}
\usepackage[colorlinks]{hyperref}
\hypersetup{
    plainpages=true,
    breaklinks=true,
    hypertexnames=false,
    pageanchor=true,
    colorlinks=true,
    linkcolor={blue},
    citecolor={red},
    urlcolor={blue},
    anchorcolor={black}
}

\usepackage[normalem]{ulem} 
\newcommand\redline{\bgroup\markoverwith
    {{\rule[.5ex]{2pt}{1pt}}}\ULon}

\begin{document}

\title{Liouvillian exceptional points of any order in dissipative
linear bosonic systems: Coherence functions and switching between
${\cal PT}$ and anti-${\cal PT}$ symmetries}

\author{Ievgen I. Arkhipov}
\email{ievgen.arkhipov@upol.cz} \affiliation{Joint Laboratory of
Optics of Palack\'y University and Institute of Physics of CAS,
Faculty of Science, Palack\'y University, 17. listopadu 12, 771 46
Olomouc, Czech Republic}

\author{Adam Miranowicz}
\email{miran@amu.edu.pl} \affiliation{Faculty of Physics, Adam
Mickiewicz University, PL-61-614 Pozna\'n, Poland}
\affiliation{Theoretical Quantum Physics Laboratory, RIKEN Cluster
for Pioneering Research, Wako-shi, Saitama 351-0198, Japan}

\author{Fabrizio Minganti}
\email{fabrizio.minganti@riken.jp} \affiliation{Theoretical
Quantum Physics Laboratory, RIKEN Cluster for Pioneering Research,
Wako-shi, Saitama 351-0198, Japan}

\author{Franco Nori}
\email{fnori@riken.jp} \affiliation{Theoretical Quantum Physics
Laboratory, RIKEN Cluster for Pioneering Research, Wako-shi,
Saitama 351-0198, Japan} \affiliation{Physics Department, The
University of Michigan, Ann Arbor, Michigan 48109-1040, USA}

\begin{abstract}
Usually, when investigating exceptional points (EPs) of an open
Markovian bosonic system, one deals with spectral degeneracies of
a non-Hermitian Hamiltonian (NHH), which can correctly describe
the system dynamics only in the semiclassical regime. A recently
proposed quantum Liouvillian framework [Minganti {\it et al.} \href{https://journals.aps.org/pra/abstract/10.1103/PhysRevA.100.062131}{Phys.~Rev.~A {\bf{100}}, 062131 (2019)}] enables the complete
determination of the dynamical properties of such systems and their EPs
(referred to as Liouvillian EPs, or LEPs) in the quantum regime by
taking into account the effects of quantum jumps, which are
ignored in the NHH formalism. Moreover, the symmetry and
eigenfrequency spectrum of the NHH become a part of much larger
Liouvillian eigenspace. As such, the EPs of an NHH form a subspace
of the LEPs. Here we show that once an NHH of a dissipative linear
bosonic system exhibits an EP of a certain finite order $n$, it
immediately implies that the corresponding LEP can become of any
higher order $m\geq n$, defined in the infinite Hilbert space.
These higher-order LEPs can be identified by the
coherence and spectral functions at the steady state. 
The coherence functions can offer a convenient tool to probe extreme
system sensitivity to external perturbations in the vicinity of
higher-order LEPs. 
As an example, we study a linear bosonic system
of a bimodal cavity with incoherent mode coupling to reveal its
higher-order LEPs; particularly, of second and third order via
first- and second-order coherence functions, respectively.
Accordingly, these LEPs can be additionally revealed by squared
and cubic Lorentzian spectral lineshapes in the power and
intensity-fluctuation spectra. Moreover, we demonstrate that these
EPs can also be associated with  spontaneous parity-time (${\cal
PT}$) and anti-${\cal PT}$-symmetry breaking in the {system} studied.
{These symmetries can be switched in the output fields (the so-called
supermodes) of an additional linear coupler with a properly chosen
coupling strength. Thus, we show that the initial loss-loss
dynamics for the supermodes can be equivalent to the balanced
gain-loss evolution.}
\end{abstract}

\date{\today}

\maketitle

\tableofcontents

\section{Introduction}
\subsection{ Exceptional points}
Exceptional points (EPs), which are exotic spectral degeneracies
of non-Hermitian open systems, have attracted much interest in
 the last decades~\cite{Ozdemir2019,Miri2019,Ashida2020}. EPs arise when both
eigenvalues and eigenfunctions of a non-Hermitian Hamiltonian
(NHH) coalesce. {Equivalently, } an NHH, at its EPs,  attains a
Jordan form, i.e., it fails to be diagonalized. The advantages of
EPs for applications remain a very active topic of
research~\cite{Wiersig2014,Zhang2015,Wiersig2016,Ren2017,Chen2017,
Hodaei2017,ChenNat2018,Liu2016,Langbein2018,
Lau2018,Mortensen2018,Zhang2018,Chen2019,Wang2020,Wiersig2020b},
in particular concerning the metrological advantages of EP
sensitivity to external perturbations~\cite{Ozdemir2019}.

The concept of EPs was first introduced in connection with the
perturbation theory of linear operators~\cite{KatoBOOK}. In
physics, the notion of the EPs was further explored in connection
with parity-time (${\cal PT}$) symmetric quantum
mechanics~\cite{Bender1998}. More recently, the concept of EPs has
been {investigated for general} open quantum systems, where the
interplay  of incoherent drives and losses with coherent coupling
can lead to the observation of an NHH spectral
degeneracy~\cite{Ozdemir2019,Miri2019,Liang2017,ChristodoulidesBook}.

The presence of EPs in a system can produce {a} plethora {of}
nontrivial phenomena. To name a few: unidirectional
invisibility~\cite{Lin2011,Regen2012},  lasers with enhanced-mode
selectivity~\cite{Feng2014,Hodaei2014}, low-power nonreciprocal
light transmission~\cite{Peng2014,Chang2014,Arkhipov2019b}, new
types of thresholdless phonon lasers~\cite{Jing2014,Lu2017},
enhanced light-matter interactions
\cite{Liu2016,Chen2017,Hodaei2017},  loss-induced lasing
\cite{Brands2014a,Peng2014a}, 
 {and even exceptional photon blockade~\cite{Huang2020}}.
 These exotic phenomena have been
observed in different experimental platforms, based on
electronics~\cite{Schindler2011}, optomechanics
\cite{Jing2014,Jing2015,Harris2016,Jing2017}, acoustics
\cite{Zhu2014,Alu2015}, plasmonics~\cite{Benisty2011}, and
metamaterials~\cite{Kang2013}. Moreover, the concept of EPs has
been successfully exploited in describing dynamical quantum phase
transitions and topological phases of condensed matter in open
quantum systems~\cite{LeykamPRL17,GonzalesPRB17,
HuPRB17,GaoPRL18,LiuPRL19,Zhou2018,BliokhNat19,
MoosSciPost19,Ge2019}, and its relation to nonclassicality in
photonic systems~\cite{Perina2019a,Perina2019b}.

\subsection{Liouvillian exceptional points}
Recently, an extension of the concept of EPs from an NHH formalism
to that based on the quantum Liouvillian has been
proposed~\cite{Minganti2019}. Indeed, the inclusion of quantum
jumps can have a profound effect on the systems dynamics and its
spectra~\cite{Prosen2012,Minganti2019,Arkhipov2020,Minganti2020,Jaramillo2020,Huber2020}.
Moreover, the eigenspectrum of an NHH becomes a part of much
larger eigenspace of a Liouvillian, meaning that EPs of an NHH,
which from hereon we denote as HEPs, become a part of
 Liouvillian EPs which are usually denoted as LEPs.
As such, the quantum Liouvillian formalism appears to be a natural
choice to study EPs of an open quantum system.
{Nonetheless, in some systems, e.g., in a double-quantum-dot circuit QED setup~\cite{Purkayastha2020}, the NHH can capture nontrivial system dynamics, which is ``invisible'' using the corresponding Liouvillian.
}

At the same time, very recent works have already shown that
switching to the Liouvillian framework allows to reveal nontrivial
phenomena, which cannot be observed within the NHH formalism.
These include  the existence of higher-order LEPs compared with
lower-order HEPs~\cite{Arkhipov2020,Wiersig2020}, or the
possibility to detect nontrivial EPs by using a hybrid Liouvillian
formalism~\cite{Minganti2020}. 
{Moreover, $\cal PT$-symmetry has been reformulated within a Liouvillian framework~\cite{Prosen2012,Huybrechts20,Huber2020}.}

\subsection{New results}
In this work, we demonstrate that for a dissipative linear bosonic
system, whose effective NHH exhibits a  HEP of any finite order
$n$, determined by first-order field moments, the corresponding
LEPs can become at least of order $m\geq 2k(n-1)+1$, for
$k=\frac{1}{2},1,2,3,\dots$, {  determined by higher-order field
moments, accordingly.} These $m$th-order LEPs can be identified by
the $(m-1)$st-order coherence  and spectral response functions in
the steady state of the system. {  Because the steady state of such systems is a thermal state, by exploiting the well-known moments theorem~\cite{AgarwalBook}, the coherence functions of arbitrary order
can be completely determined by the first-order coherence
function, i.e., by its products.
As such, by identifying a HEP with the first-order
coherence function~\cite{Arkhipov2020}, one then can identify
higher-order LEPs by means of the corresponding higher-order coherence functions.
Importantly, the coherence functions can only be determined by the {\it Liouvillian eigenspace}.
}

We stress that when considering an NHH in an infinite-dimensional
system, one can also determine infinitely-high-order HEPs related
to high-order moments of the fields~\cite{Juarez2019,Zhang2020b}.
Nevertheless, NHHs fail to include quantum jumps and thermal
noise. A striking example of this fact comes from an effective
Hamiltonian which commutes with the total photon number operator.
While at the NHH level this implies that manifolds with different
numbers of excitations cannot interact, the quantum jumps of the
Liouvillian can still mix  states with different photon
numbers. These different properties of the NHH and the
corresponding Liouvillian imply that also their eigenstates are
different~\cite{Minganti2019,Arkhipov2020}. As a result, the
equations of motion for second- or higher-order quantum field
moments differ for the Liouvillian and NHH formalisms, indicating
their distinguishable spectral properties. Hence, the analyses of
higher-order HEPs resulting from higher-order moments are
predictive only in the semiclassical regime, when the operators
can be treated as $c$-numbers. In other words, {\it a correct
description of spectral properties of a quantum system via its
higher-order field moments needs to rely exclusively on the
Liouvillian eigenspectrum.}

\subsection{Higher-order Liouvillian exceptional points in a bimodal cavity with
incoherent mode coupling}
As an example of the above general result, we study a bosonic
system of a bimodal cavity with incoherent mode coupling to reveal
its higher-order Liouvillian EPs. The incoherent mode interaction
can be encoded by the off-diagonal elements of the damping matrix
in the quantum Liouvillian~\cite{Haake2003,Franke19}. These
off-diagonal damping coefficients naturally appear in the
microscopic theory of overlapping modes in open
resonators~\cite{Haake2003} and chaotic two-mode
lasers~\cite{Eremeev2011}, where a strong interaction with the
surrounding environment can induce a mode overlapping in the
multimode cavities. Moreover, these theories have proved useful in
explaining the Petermann excess noise factor in random
lasers~\cite{Eremeev2011,Grangier1998} and describing
intensity-fluctuation spectra in bimodal cavities coupled to
quantum emitters~\cite{Wiersig2013,Wiersig2016b}. Interestingly,
one of the first experimental observations of the intermode
coupling in a bimodal cavity due to the interaction with
surrounding screening fields, induced by a conducting sample
located near the cavity, was already reported a few decades ago in
Ref.~\cite{Ong1981}, where microwave Hall measurements were
performed. Incoherent mode coupling can be produced in various
ways. These have been already realized, e.g.,  in
anti-$\cal PT$-symmetrical {\it classical} systems, which include:
nonlinear Brillouin scattering in a single
microcavity~\cite{Zhang2020}, two passive waveguides, separated by
a metallic film~\cite{Fan2020}, countermoving media with heat
exchange~\cite{Li2019} or resistively coupled electric
resonators~\cite{Choi2018}; and, in quantum systems, through a
coherent transport of flying atoms~\cite{Peng2016}. Nevertheless,
in all these works, when studying EPs, a phenomenological approach
has been utilized based exclusively on effective NHHs, thus
ignoring quantum-jump effects.

In particular, we analyze second and third-order LEPs of such
systems arising from a HEP of  second order by calculating the
first- and second-order coherence functions, respectively. Also,
we calculate the corresponding power and intensity-fluctuation
spectra to reveal their squared and cubic Lorentzian expressions,
accordingly. We also reveal the anti-$\cal PT$- and $\cal PT$-symmetries of
the NHH, which connect the presence of LEPs with spontaneous
breaking  of such symmetries in a bimodal cavity with incoherent
mode coupling. We show the possibility of switching between the
\PT and anti-\PT symmetries of bosonic linear systems, like
linearly-coupled harmonic cavities. This switching can simply be
realized by applying a tunable linear coupler (e.g., a tunable
beam splitter in optical implementations) to a two-mode output
field of the system. 
{Moreover, our analysis of such two-mode systems reveals that for an LEP of odd order $(2k+1)$,
the system enhanced sensitivity to external perturbations $\epsilon$ scales at most as $\epsilon^{\frac{1}{2k}}$.}

{We stress that the  discussed \PT and anti-\PT
	symmetries in our study are exclusively related to the symmetries
	of the NHH, which plays a central role in our work. However, we
	note that the recent studies in Refs.~\cite{Prosen2012,Huber2020,Huybrechts20} have already
	addressed the properties of $\cal PT$-symmetry of the whole Liouvillian.
	Since here we focus on dissipative systems, the whole Liouvillian
	does not possess the $\cal PT$-symmetry in the sense of, e.g.,
	Ref.~\cite{Bender1998}. Nonetheless, the studied Liouvillian, for a
	two-site system, is {\it passively} $\cal PT$-symmetric. In other words,
	it acquires the $\cal PT$-symmetry in a reference frame with global
	decay~\cite{Prosen2012,Huber2020}; that is, after applying an appropriate gauge
	transformations.
} 

Let us recall now the meaning of the \PT and anti-\PT
symmetries of the NHH. In general, a system described by a Hamiltonian $\hat
H$ exhibits the \PT (anti-\PT) symmetry if $\hat H$ commutes
(anticommutes) with the \PT operator. Here the parity operator
${\cal P}$ transforms a position operator $\hat x$ to $-\hat x$
and a momentum operator $\hat p$ to $-\hat p$, while the time
reversal operator ${\cal T}$ transforms $\hat x\rightarrow \hat x$
and $\hat p\rightarrow -\hat p$, and performs complex conjugation
$i \rightarrow -i$.

This paper is organized as follows. In Sec.~\ref{II}, we briefly
introduce a general Liouvillian for a dissipative multimode
bosonic system. In Sec.~\ref{III}, we present our main result,
namely that any HEP of a finite order implies the infinite order
of the corresponding LEP. As an example, we study higher-order
LEPs in a bimodal cavity with incoherent mode coupling in
Sec.~\ref{IV}. In particular,  we analyze second and third-order
LEPs by means of the first and second-order coherence functions,
respectively, along with the power and intensity-fluctuation
spectra to reveal their squared and cubic Lorentzian lineshapes at
the corresponding LEPs. We also demonstrate that EPs in such
systems can be directly associated with \PT and anti-$\cal PT$-symmetry
breaking. Conclusions are drawn in Sec.~\ref{V}.


\section{Liouvillian of a general dissipative linear bosonic system}\label{II}
The dynamics of a density matrix $\hat\rho$ describing a {quantum
system interacting with its environment}  is governed by {a
completely positive trace-preserving (CPTP) map.} {In the limit of
weak Markovian time-independent interactions, such a CPTP map is
known as the}  Liouvillian superoperator $\cal L$ {whose action is
described } by the master equation:
\begin{equation} \label{ME} 
\frac{{\rm d}}{{\rm d}t}\hat\rho(t)={\cal L}\hat\rho(t).
\end{equation}
For an $N$-mode open linear coupled bosonic system interacting
with thermal environment, the Liouvillian has  the following general
Gorini-Kossakowski-Sudarshan-Lindblad form
($\hbar=1$)~\cite{Haake2003,Franke19}:
\begin{eqnarray} \label{L} 
{\cal L}\hat\rho=&&-i\left[\hat H,\hat\rho\right]+\frac{n_{\rm th}+1}{2}\sum\limits_{j,k}^N\gamma_{jk}{{\cal D}}\left[\hat a_{j},\hat a_k^{\dagger}\right]\hat\rho \nonumber \\
&&+\frac{n_{\rm th}}{2}\sum\limits_{j,k}^N\gamma_{jk}{\cal
D}\left[\hat a_{j}^{\dagger},\hat a_k\right]\hat\rho,
\end{eqnarray}
where $\hat H$ is a Hermitian Hamiltonian, and the general
Lindblad dissipators  are
\begin{equation}\label{LD} 
{{\cal
D}}\left[\hat\Gamma_{j},\hat\Gamma_k^{\dagger}\right]\hat\rho=2\hat\Gamma_{j}\hat\rho\hat
\Gamma_{k}^{\dagger}-\hat \Gamma_{k}^{\dagger}\hat
\Gamma_{j}\hat\rho-\hat\rho\Gamma_{k}^{\dagger}\hat \Gamma_{j}.
\end{equation}
In Eq.~(\ref{L}), $\hat a_j$ ($\hat a_j^{\dagger}$) is the
annihilation (creation) operator of  mode $j$; the diagonal
damping coefficients $\gamma_{kk}$ denote the inner $k$th mode
{decay rate}, while the off-diagonal coefficients $\gamma_{jk}$
denote the {\it incoherent} coupling between modes $j$ and $k$,
due to the interaction of both modes with the
environment~\cite{Haake2003}. Without loss of generality, we
assume that the thermal photon number $n_{\rm th}$ is constant
throughout the spectral range of a system. The Liouvillian $\cal
L$ can also be recast in the following form
\begin{eqnarray}\label{eq4} 
{\cal L}\hat\rho=&&-i\left(\hat H_{\rm eff}\hat\rho-\hat\rho\hat H^{\dagger}_{\rm eff}\right)+\frac{n_{\rm th}+1}{2}\sum\limits_{j,k}\hat a_{j}\hat\rho\hat a_{k}^{\dagger}\nonumber \\
&&+\frac{n_{\rm th}}{2}\sum\limits_{j,k}\hat
a_{j}^{\dagger}\hat\rho\hat a_{k},
\end{eqnarray}
where $\Hef$ is an effective NHH given by:
\begin{equation}\label{Heff} 
\Hef=\hat H-\frac{i}{2}\sum\limits_{j,k}\gamma_{jk}\hat
a_j^{\dagger}\hat a_k.
\end{equation}
 Note that the term $\left(\hat H_{\rm eff}\hat\rho-\hat\rho\hat
H^{\dagger}_{\rm eff}\right)$ in Eq.~(\ref{eq4}) can be
interpreted as a generalized commutator. Moreover, this NHH is not
Hermitian, i.e, $\Hef\neq\hat H_{\rm eff}^{\dagger}$.
Additionally, the Hermitian Hamiltonian in Eq.~(\ref{Heff}) for a
linear coupled system can be written in a general form
\begin{equation}\label{Hc} 
\hat H=\sum\limits_k\omega_{k}\hat a^{\dagger}_k\hat
a_k+\sum\limits_{j<k}\left(\chi_{jk}\hat a_{j}^{\dagger}\hat
a_k+{\rm H.c.}\right),
\end{equation}
where $\omega_k$ is a bare frequency of the mode $k$, and
$\chi_{jk}$ is the {coherent} coupling coefficient between modes
$j$ and $k$.

\section{Liouvillian exceptional points of any order in dissipative linear bosonic systems}\label{III}

In a recent work~\cite{Wiersig2020}, it has been shown that if  a
NHH has an EP of  $n$th order, then it implies that a LEP is at
least of   order $(2n-1)$. Below, we demonstrate in a simple
manner that if   an NHH of a linear bosonic system has an EP of any
order $n\geq2$, determined by the first-order field moments, then
this EP actually implies  an infinite order for the LEP, which,
accordingly, is determined by higher-order field moments, and,
thus, can be identified by higher-order coherence functions.

To show this general result, let us first start  describing
the time dynamics of the system field averages $\langle\hat
a_{j}(t)\rangle$. Note that we are working exclusively in the
Schr\"{o}dinger picture;  thus, for simplicity, we put the time
parameter $t$ inside  triangular brackets. After applying the formula
for  the time derivative of the field averages
\begin{equation}\label{TR} 
\frac{{\rm d}}{{\rm d}t}\langle\hat a_{j}(t)\rangle={\rm
Tr}\left[\hat a_j\frac{{\rm d}}{{\rm d}t}\hat\rho(t)\right],
\end{equation}
and using Eqs.~(\ref{ME}), (\ref{L}), and (\ref{TR}), one obtains
a linear system
\begin{equation}\label{RES} 
v(t)=\exp(-i\hat H_{\rm eff}t) v(0)
\end{equation}
where $v(t)=[\langle\hat a_1(t)\rangle,\langle\hat
a_2(t)\rangle,\dots,\langle\hat a_N(t)\rangle]^T$ is a vector of
the operator averages, and $\Hef$ is a matrix form of the
effective NHH in Eq.~(\ref{Heff}).

Hence, the dynamics of the annihilation operators imposed by the
Liouvillian $\cal L$ can be fully determined by the eigenspectrum
of the effective NHH $\Hef$. As a result, the appearance of an EP
in the NHH spectrum, immediately implies the emergence of the same
EP in the Liouvillian spectrum~\cite{Arkhipov2020}. Namely, the
relationship between the eigenfrequencies (eigenvalues):
$\nu\equiv\nu_{\rm NHH}$ of the NHH and
$\lambda\equiv\lambda_{\cal L}$ of the Liouvillian, which define
the time dynamics of the fields in Eq.~(\ref{RES}), bear a simple
form~\cite{Arkhipov2020}:
\begin{equation}\label{eq9} 
\lambda_{\cal L}=-i\nu_{\rm NHH}.
\end{equation}
Therefore, the coalescence of the eigenvalues $\nu_{\rm NHH}$ of
the NHH, along with its eigenstates, indicates the coalescence of
the eigenvalues $\lambda_{\cal L}$ and the corresponding
eigenstates of the  Liouvillian.  These eigenvalues merging  cause
the NHH to acquire a nondiagonal Jordan form;  which is, then,
reflected in the nonexponential character  of the time evolution of
the cavity fields, according to Eq.~(\ref{RES}). Moreover, the
symmetry shared by the NHH becomes, in general, a local symmetry
of the Liouvillian. The latter stems from the fact that the
Liouvillian does not necessarily have  the global symmetries of the
NHH.
{We stress that the above conclusion of the
	coincidence of EPs of the NHH and Liouvillian is valid for any
	linear quadratic NHH in  Eq.~(\ref{Heff}), with its coherent part given in
	Eq.~(6). Moreover, the damping coefficients in the Lindblad
	dissipators in Eq.~(\ref{LD}) may attain any form, e.g., similar to that
	of the Scully-Lamb laser model~\cite{Arkhipov2020}, as long as the system remains
	linear, i.e., dissipative.
}

\subsection{ Higher-order correlation functions}
The exact equivalence between the effective Hamiltonian and
Liouvillian predictions of the spectral properties of the system
holds true only for the dynamics of the annihilation operators.
For example, higher-power field averages $\langle\hat
a_{j}^{\dagger m}\hat a_j^n(t)\rangle$ would be affected by the
presence of quantum jumps in a nontrivial way. Nevertheless, from
the presence of a first-order HEP one we can deduce the properties
of higher-order correlation functions, which, for linear systems,
are determined by the higher-order field moments. Indeed, this
function, at the steady state,  can be calculated according to the
formula:
 \begin{equation}\label{g1def} 
g^{(1)}_{j,{\rm ss}}(\tau)=\frac{\langle\hat a_j^{\dagger}(0)\hat
a_j(\tau)\rangle_{\rm ss}}{\langle\hat a_j^{\dagger}(0)\hat
a_j(0)\rangle_{\rm ss}}, \quad j=1,\dots,N,
\end{equation}
where $\langle\hat a_j^{\dagger}(0)\hat a_j(\tau)\rangle_{\rm ss}$
is a two-time correlation function (TTCF) for the mode $j$ at the
steady state. The TTCF, in turn, can be easily computed by
exploiting the quantum regression
theorem~\cite{CarmichaelBook,AgarwalBook}. Namely, by solving the
equations of motion for the field averages $\langle \hat
a(\tau)\rangle$ in Eq.~(\ref{RES}), one can immediately obtain
$\langle\hat a^{\dagger}(0)\hat a(\tau)\rangle$. The TTCF for,
e.g., the field $\hat a_1$ reads as~\cite{Arkhipov2020}:
\begin{equation}\label{RES1} 
f_j(\tau)=\exp(-i\hat H_{\rm eff}\tau) f_j(0),
\end{equation}
where $f(\tau)=[\langle\hat a_j^{\dagger}(0)\hat
a_1(\tau)\rangle_{\rm ss},\dots,\langle\hat a_j^{\dagger}(0)\hat
a_N(\tau)\rangle_{\rm ss}]^T$ is a vector of TTCFs for the mode
$j$.

 In other words, the dynamics and symmetry of the equations of
motion for the coherence function $g^{(1)}{(\tau)}$ is determined
by the same effective NHH. Again, at the HEP of $n$th order, the
NHH obtains a Jordan form. As a result, and according to
Eq.~(\ref{RES1}), the coherence function $g^{(1)}{(\tau)}$,
regardless of the mode $j$, attains a nonexponential form, with
the highest power degree in $\tau$, as follows:
\begin{equation}\label{g1t} 
 g^{(1)}_{\rm ss}(\tau)\sim\tau^{n-1}\exp\left(-i\nu_{\rm HEP}\tau\right),
\end{equation}
where $\nu_{\rm HEP}$, with imaginary part ${\rm Im}(\nu_{\rm
HEP})<0$, is a complex eigenvalue of the NHH at an HEP of $n$th
order.  

 The steady state of the considered linear systems with the
Liouvillian $\cal L$, given in Eq.~(\ref{L}), along with
 the Hamiltonian in Eq.~(\ref{Hc}), is a thermal state. As a result,
the higher-order coherence functions $g^{(k)}_{\rm ss}(\tau)$,
$k\in {\mathbb N}$, at the steady state, are completely determined
by the first-order coherence function $g^{(1)}_{\rm ss}(\tau)$,
according to the moments theorem~\cite{AgarwalBook}. 

 The
higher-order coherence functions $g^{(2k)}_{j,\rm ss}(\tau)$,
based on the TTCFs, are found as
\begin{equation}\label{g2def} 
g^{(2k)}_{j,\rm ss}(\tau)=\frac{\langle \hat a_j^{\dagger
k}(0)\hat a_j^{\dagger k}(\tau)\hat a_j^{k}(\tau)\hat
a_j^{k}(0)\rangle_{\rm ss}}{\langle\hat a_j^{\dagger}(0)\hat
a_j(0)\rangle_{\rm ss}^{2k}}.
\end{equation}
The form of the coherence function $g^{(2k)}_{j,\rm ss}(\tau)$ in
Eq.~(\ref{g2def}) ensures that
 at the $n$th-order HEP, the $(2k)$th-order coherence function at the steady state contains the following term with the highest  power degree in $\tau$:
  \begin{equation}\label{g2k} 
 g^{(2k)}_{\rm ss}(\tau)\sim\tau^{2k(n-1)}\exp\Big[2k{\rm Im}(\nu_{\rm HEP})\tau\Big],
 \end{equation}
For instance,  the second-order coherence function $g^{(2)}_{\rm
ss}(\tau)$ for the thermal light takes a simple form
 \begin{equation}\label{g2g1} 
g^{(2)}_{\rm ss}(\tau)=1+\left|g^{(1)}_{\rm
ss}(\tau)\right|^2\sim\tau^{2n-2}\exp\Big[2{\rm Im}\left(\nu_{\rm
HEP}\right)\tau\Big].
\end{equation}

 The coherence function $g^{(2k)}_{\rm ss}(\tau)$ is solely
defined by the Liouvillian eigenspectrum;  thus, at the HEP of the
$n$th order, this function implies the coalescence of
$[2k(n-1)+1]$ Liouvillian eigenvectors. This means that  the LEP
becomes at least of the order $[2k(n-1)+1]$. Moreover, because of
the infinite-dimensional Hilbert space of a general bosonic
system, the coherence function can be, thus, of  infinite
order.  Thus, a HEP  would simply imply the existence of the LEP
of {\it infinite} order. In other words, the order of an LEP is
only limited by the maximal possible number of photons in a
system, i.e., the maximal size of its Hilbert space.

To shed more light on this direct connection between the
higher-order LEPs and higher-order coherence functions, let us
recall the general formula for the TTCFs for the steady state,
which is used in the definition of the coherence function:
\begin{equation}\label{O123} 
\langle \hat O_1(0) \hat O_2(\tau)\hat O_3(0)\rangle_{\rm ss}={\rm
Tr} \left\{\hat O_2(0) e^{\mathcal{L} \tau} \left[\hat
O_3(0)\hat\rho_{\rm ss} \hat O_1(0)\right] \right\},
\end{equation}
where $\hat O_j$ are some system operators. The  operator $\hat
O_3\hat{\rho}_{\rm ss}\hat O_1$, in Eq.~(\ref{O123}), leads the
steady state $\hat\rho_{\rm ss}$ into the new state, which becomes
a decomposition of the Liouvillian eigenmatrices $\hat\rho_i$,
i.e.,
\begin{equation}\label{rhoss}  
\hat O_3\hat{\rho}_{\rm ss}\hat O_1 = \sum_i c_i \hat{\rho}_i.
\end{equation}
By recalling the linearity of the trace, we have
\begin{equation} \label{TTCFdef} 
\langle \hat O_1(0) \hat O_2(\tau)\hat O_3(0)\rangle_{\rm
ss}=\sum_i c_i  {\rm Tr} \left\{\hat O_2(0) e^{\mathcal{L} \tau}
\left[\hat{\rho}_i \right] \right\}.
\end{equation}
In the presence of an LEP, one finally has
\begin{equation}\label{TTCFEP} 
\langle \hat O_1(0) \hat O_2(\tau)\hat O_3(0)\rangle_{\rm
ss}=\sum_i c_i \tau^{n_i} e^{\lambda_i \tau}  {\rm Tr} \left\{\hat
O_2(0) \hat{\rho}_i \right\},
\end{equation}
where $n_i$ is the degree of the degeneracy of the LEP associated
with the eigenmatrix $\hat{\rho}_i$ (see also
Ref.~\cite{Arkhipov2020} for more details). Moreover, the
eigenmatrices $\hat\rho_i$ correspond to various powers of the
boson operators of the fields~\cite{Honda2010}, which means that
this is the eigenspace of higher-order field moments that is
identified by the TTCFs. In our particular case, $\hat O_1=\hat
a^{\dagger k}$, $\hat O_2=\hat a^{\dagger k}\hat a^{k}$, and $\hat
O_3=\hat a^{k}$. Note that TTCFs, in Eq.~(\ref{O123}), can be
applied to any moments of the field, i.e., not necessarily to the
Hermitian moments $O_2(\tau)=\hat a^{\dagger k}(\tau)\hat
a^{k}(\tau)$. As such, it is possible to reveal an arbitrary order
$m$ of a LEP, apart from that identified by the coherence function
$g^{(2k)}_{\rm ss}(\tau)$.

Experimentally, the spectral properties of dissipative systems
have been measured, in particular regarding the closure of the
Liouvillian gap occurring in dissipative phase
transitions~\cite{Minganti2018}. In particular, in
Ref.~\cite{FinkNatPhys18} the two-time correlation function has
been used to prove the occurrence of the first-order phase
transition of a semiconductor micropillar, as predicted in
Refs.~\cite{Bartolo2016, Casteels2017}. Moreover, optical
hysteresis properties have been used in Ref.~\cite{RodriguezPRL17}
to prove the emergence of a critical slowing-down effect.
Similarly, in Ref.~\cite{FitzpatrickPRX17} the emergence of a slow
timescale in a one-dimensional superconductor chain has been used
to pinpoint the precursors of a dissipative phase transition in a
driven-dissipative Bose-Hubbard
model~\cite{Foss-Feig2017,Vicentini2018}.

We conclude that observing LEPs using the first- and second-order
correlation functions is within the experimental reach of current
techniques. For instance, an $n$th-order HEP would imply the
possibility to measure a LEP at least of order
$(2n-1)$~\cite{Wiersig2020} by means of the coherence function
$g^{(2)}_{\rm ss}(\tau)$. We note that, however, to access
higher-order LEPs using higher-order coherence functions could be
much more  challenging. Indeed, $g^{(2k)}_{\rm ss}(\tau)$ decays
much faster than $g^{(2)}_{\rm ss}(\tau)$ [c.f. Eq.~(\ref{g2k})],
thus reducing the visibility of its nonexponential behavior.

\section{Example of a bimodal cavity with incoherent mode coupling}\label{IV}

In this section, we study higher-order LEPs in a bimodal cavity
with incoherent mode coupling. In particular, we analyze second-
and third-order LEPs arising from a HEP of the second order by
calculating first- and second-order coherence functions,
respectively. Also, we calculate the corresponding power and
intensity-fluctuation spectra to reveal their squared and cubic
Lorentzian lineshapes, accordingly. Additionally, we reveal the
anti-$\cal PT$- and $\cal PT$-symmetries of the NHH, which connect the
presence of the LEPs with spontaneous breaking  of such symmetries
in the system.

\subsection{Non-Hermitian Hamiltonian exceptional second-oder points and its anti-$\cal PT$- and $\cal PT$-symmetries}

\subsubsection{Anti-$\cal PT$-symmetry and exceptional point of an effective non-Hermitian Hamiltonian}

The dynamics of a density matrix $\hat\rho$ of a bimodal cavity
with incoherent mode coupling is described by the Liouvillian in
Eq.~(\ref{L}) with the free coherent Hamiltonian of the form
\begin{equation}\label{Hfree} 
\hat H=\sum\limits_{k=1,2}\omega_k\hat a^{\dagger}_k\hat a_k.
\end{equation}
For simplicity, we further assume that the damping matrix
$\gamma_{jk}$ is symmetric, i.e., $\gamma_{21}=\gamma_{12}$, and
the inner mode decaying rates are the same, i.e.,
$\gamma_{11}=\gamma_{22}=\gamma$.

By working in the rotating reference frame with the central
frequency $\bar\omega=(\omega_1+\omega_2)/2$, the effective NHH,
given in Eq.~(\ref{Heff}), attains the form
\begin{equation} \label{HeffMa} 
\hat H_{\rm eff}=\frac{1}{2}\begin{pmatrix}
\Delta-i{\gamma} & -i\gamma_{12} \\
-i\gamma_{12} & -\Delta-i\gamma
\end{pmatrix},
\end{equation}
where $\Delta=(\omega_1-\omega_2)$ is a cavity resonance
difference.

This NHH is anti-$\cal PT$-symmetric, meaning that its anticommutator
with a \PT operator is zero:
\begin{equation} 
{\cal PT}\Hef({\cal PT})^{-1}=-\Hef.
\end{equation}
The action of the time-reversal operator $\cal T$ on the NHH
$\Hef$ is equivalent to
\begin{equation*}
{\cal T}\Hef{\cal T}^{-1}={\Hef}^*,
\end{equation*}
 where the asterisk indicates the complex conjugation.
And the parity operator $\cal P$ is equivalent to the Pauli
$\hat\sigma_x$ matrix. The presence of the incoherent mode
coupling rate $\gamma_{12}$, thus, induces the anti-$\cal PT$-symmetry
in the evolution of the field averages.

\begin{figure} 
\includegraphics[width=0.4\textwidth]{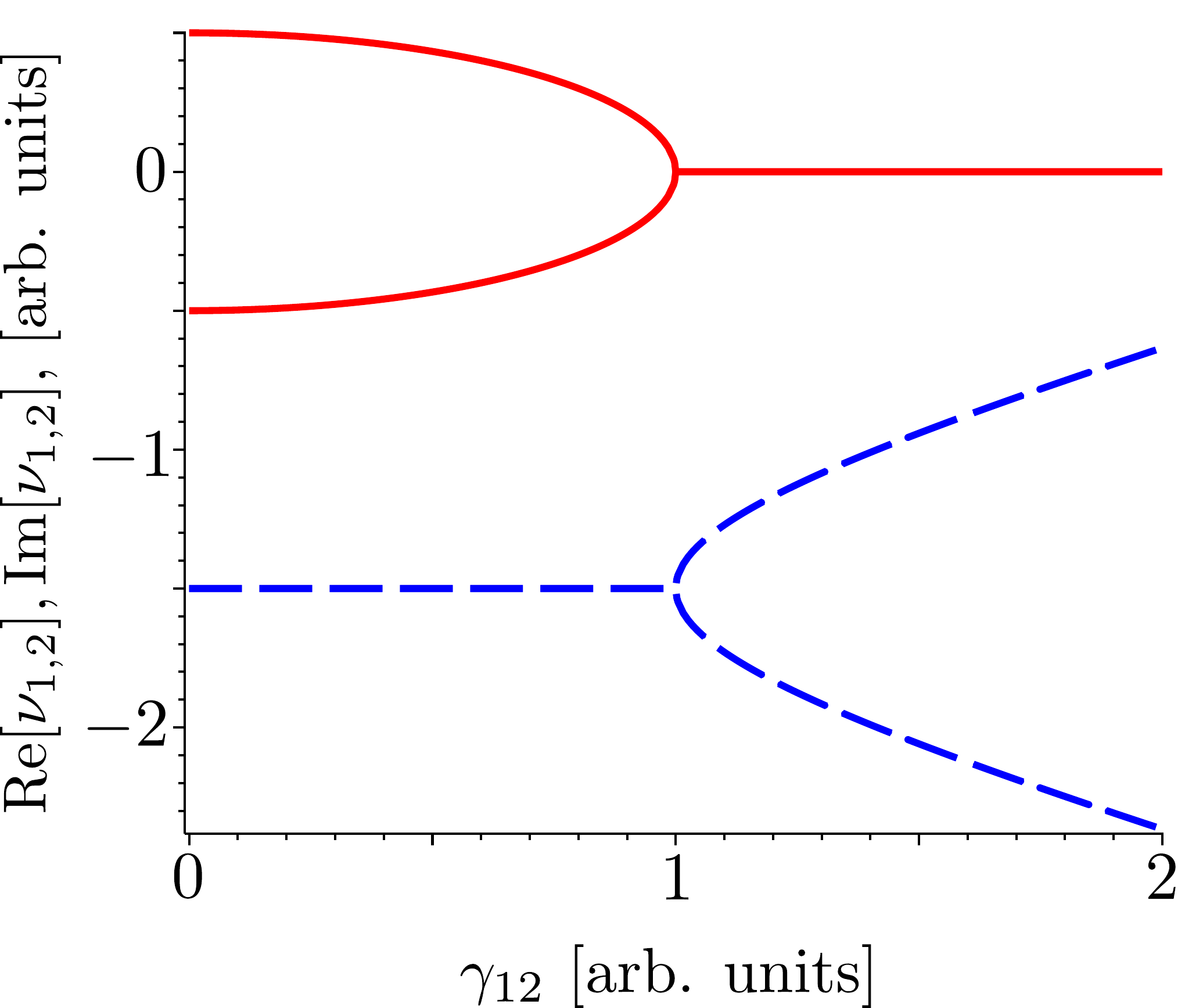}
\caption{Real (red solid curves) and imaginary (blue dashed
curves) parts of the eigenfrequencies $\nu_{1}$ and $\nu_2$ of the effective
NHH $\hat H_{\rm eff}$ versus the incoherent coupling rate
$\gamma_{12}$. These eigenfrequencies constitute a subset of the
eigenfrequencies of the Liouvillian $\cal L$, of a bimodal cavity
according to Eq.~(\ref{nu12}). The chosen system parameters are:
the frequencies of the modes $\omega_2=-\omega_1=0.5$ [arbitrary
units]; the cavity losses for both modes $\hat a_1$ and $\hat a_2$
are $\gamma=3$ [arb. units]. The system experiences a
spectral-phase transition at the EP $\gamma_{12}^{\rm EP_1}$,
according to Eq.~(\ref{HEP}), due to the interplay between the
mode frequency difference $\Delta$ and the incoherent mode
coupling $\gamma_{12}$.}\label{fig1}
\end{figure}

The eigenspectrum of the NHH gives two possible eigenvalues:
\begin{equation} \label{nu12} 
\nu_{1,2}=-i\frac{\gamma}{2}\pm\frac{i}{2}D,
\end{equation}
where
\begin{equation} \label{D} 
D=\sqrt{\gamma_{12}^2-\Delta^2}.
\end{equation}
The unnormalized eigenvectors of the NHH, then, can be easily
found as
\begin{equation} \label{psi12} 
\psi_{1,2}\equiv
\begin{pmatrix}
i\gamma_{12} \\
\Delta\mp iD
\end{pmatrix}.
\end{equation}
From Eqs.~(\ref{nu12}) and (\ref{psi12}) it follows that the NHH
$\Hef$, and, thus, the Liouvillian, attain an EP
\begin{equation} \label{HEP}  
\gamma_{12}^{\rm EP}=|\Delta|=|\omega_1-\omega_2|.
\end{equation}
At this EP, the NHH experiences a  spectral-phase transition,
associated with anti-$\cal PT$-symmetry breaking. Meaning that cavity
fields in Eq.~(\ref{RES}) can exist in two different spectral
phases.

We plot  the eigenfrequencies $\nu_{1}$ and $\nu_2$ in Fig.~\ref{fig1}. When
$\gamma_{12}>\Delta$, i.e., the NHH eigenvalues $\nu_{1,2}$  both
become purely imaginary, the NHH eigenstates $\psi_{1}$ and $\psi_2$ are in
the exact anti-\PT symmetric phase. That is, the eigenmodes
$\psi_1$ and $\psi_2$ become asymmetric, the former attaining an
effective gain and the latter  acquiring additional losses
proportional to  the parameter $|D|$ [see Eq.~(\ref{nu12})]. Contrary
to this, when  $\gamma_{12}<\Delta$, the eigenvalues are no longer
purely imaginary, and the eigenstates of the NHH are in the broken
anti-$\cal PT$-symmetric phase. In this case, the two modes are
spectrally separated, and this mode splitting can be defined by
the real-valued parameter $|D|$ given in  Eq.~(\ref{nu12}).


\subsubsection{Switching to the $\cal PT$-symmetric modes}
By appropriate unitary transformations, the effective NHH in
Eq.~(\ref{HeffMa}) can be recast into a form, where it acquires a
$\cal PT$-like symmetry, namely, {\it passive} $\cal PT$-symmetry. For this,
we can introduce the general \emph{combined modes} (often referred
to as \emph{supermodes}) $\hat c_1$ and $\hat c_2$ (see,
e.g.,~\cite{Svozil1990,Wiersig2013}), defined via the rotation
\begin{equation}\label{c1c2} 
\begin{pmatrix}
\hat c_1 \\
\hat c_2
\end{pmatrix}=
\begin{pmatrix}
\cos\theta & -\sin\theta \\
\sin\theta & \cos\theta
\end{pmatrix}
\begin{pmatrix}
\hat a_1 \\
\hat a_2
\end{pmatrix},
\end{equation}
where $\theta$ is an appropriate angle. This transformation of
$\hat a_1$ and $\hat a_2$ to the supermodes $\hat c_1$ and $\hat
c_2$ can simply be realized with a tunable linear coupler. In case
of optical implementations of our system, this coupler can be
realized by a single tunable beam splitter or, in more refined
implementations, by a Mach-Zehnder
interferometer~\cite{MandelBook}.

For the case under consideration, i.e., $\gamma_{12}=\gamma_{21}$,
the Lindblad master equation, given in \eqref{L}, can be put in
the diagonal form of the damping matrix $\gamma_{jk}$ by
considering $\theta=\pi/4$, i.e., $\hat{c}_{1,\, 2}=(\hat a_1 \pm
\hat a_2)/\sqrt{2}$. We have:
\begin{equation} 
\begin{split}
{\cal L} \hat\rho = -i[\hat H_{c_1,c_2},\hat\rho]+ &\sum_{k=1, \,
2} \frac{\gamma_{c_k}(n_{\rm th}+1)}{2} {\cal D} [\hat c_k] \rho
\\ & + \frac{\gamma_{c_k} n_{\rm th} }{2} {\cal D} [\hat
c_k^\dagger] \rho,
\end{split}
\end{equation}
where
\begin{equation}\label{Hc1c2} 
\hat H_{c_1,c_2}=\sum\limits_k\bar\omega\hat c_k^{\dagger}\hat
c_k+\frac{\Delta}{2}\left(\hat c_1^{\dagger}\hat c_2+\hat
c_2^{\dagger}\hat c_1\right),
\end{equation}
where $\gamma_{c_1,c_2}=\gamma\mp\gamma_{12}$. Hence, the model of
the two incoherently coupled  modes $\hat a_{1}$ and$\hat a_{2}$ becomes that of
two dissipative coherently coupled modes $\hat c_{1}$ and $\hat c_{2}$ in the
appropriate basis (see Appendix~\ref{AA}) for a general form of
the  Liouvillian $\cal L$, under the transformation in
Eq.~(\ref{c1c2}).

The effective NHH $\Hef$ in Eq.~(\ref{HeffMa}), in a rotating
reference frame $\bar\omega$, then reads
\begin{equation}\label{Hc1c2ef} 
\hat H_{c_1,c_2}^{\rm eff}=\frac{1}{2}
\begin{pmatrix}
-i\gamma_{c_1} & {\Delta} \\
{\Delta} & -i\gamma_{c_2}
\end{pmatrix}.
\end{equation}

This effective NHH in Eq.~(\ref{Hc1c2ef}) now indicates that the
supermodes $\hat c_{1}$ and $\hat c_{2}$ constitute a {\it passive} \PT-symmetric
system~\cite{Ozdemir2019,Miri2019,Liang2017,ChristodoulidesBook}.
Namely, if one applies a gauge transformation
\begin{equation}\label{gauge} 
\begin{pmatrix}
\hat c_1^{'} \\
\hat c_2^{'}
\end{pmatrix}=\exp{\left(-\frac{\gamma}{2} t\right)}
\begin{pmatrix}
\hat c_1 \\
\hat c_2
\end{pmatrix},
\end{equation}
the modified NHH in Eq.~(\ref{Hc1c2ef}) then reads
\begin{equation}\label{Hc1c2efprime} 
\hat H_{c_1,c_2}^{'{\rm eff}}=\frac{1}{2}
\begin{pmatrix}
i\gamma_{12} & {\Delta} \\
{\Delta} &-i\gamma_{12}
\end{pmatrix}.
\end{equation}
This NHH $\hat H_{c_1,c_2}^{'{\rm eff}}$, in
Eq.~(\ref{Hc1c2efprime}), commutes with the \PT operator, i.e.,
\begin{equation}\label{comm} 
\left[\hat H_{c_1,c_2}^{'{\rm eff}},{\cal PT}\right]=0.
\end{equation}
In other words,   {\it the initial loss-loss dynamics} for the supermodes
$\hat c_{1}$ and $\hat c_{2}$  {\it  becomes equivalent to the balanced gain-loss
evolution}, apart from the global decay rate $\gamma/2$.

\subsubsection{ Summary}
To sum up, we have shown that, by applying appropriate unitary
transformations to the anti-$\cal PT$-symmetric NHH in Eq.~(\ref{Heff}),
one can readily discover a hidden $\cal PT$-symmetry of the NHH. As
such, the EP in Eq.~(\ref{HEP}) is associated not only with the
anti-$\cal PT$-symmetry but also with $\cal PT$-symmetry breaking of the NHH,
induced by the same interplay between the frequency difference
$\Delta$ and incoherent coupling rate $\gamma_{12}$.

Note that the exact (broken) anti-$\cal PT$-symmetric phase is
accompanied by the broken (exact) $\cal PT$-symmetric phase. Indeed, the
eigenfrequencies of the NHH are left unchanged under the unitary
transformations in Eq.~(\ref{c1c2}), but the action of the \PT and
anti-\PT symmetry is opposite for  the purely imaginary
eigenfrequencies. Such coexistence of the opposite symmetric
phases has already been pointed out in Refs.~\cite{Ge2017,Zhang2020}.
We stress that the possibility to witness different symmetries of
the system by considering different operators offers  great
flexibility to explore different dynamical regimes and various
kinds of nontrivial light behavior in this system in the
semiclassical regime~\cite{Li2019,Zhang2020,Fan2020}. Moreover, we
have explained that one can witness not only the \PT and anti-\PT
symmetries, but can also physically switch between them by
transforming the system two-mode output fields with an additional
tunable linear coupler, e.g., a tunable beam splitter in optical
implementations of the general model  discussed.

\subsection{Liouvillian exceptional points of higher orders}
As it has been already stressed in Sec.~\ref{III}, {\it an HEP of any
order $n$ implies an LEP of any higher order $m\geq n$}. Below, we study the second and
third order LEPs, which arise due to the presence of a
second-order HEP in the system  under consideration.
\subsubsection{LEP of second order and squared Lorentzian power spectra}
\begin{figure}[t!] 
\includegraphics[width=0.48\textwidth]{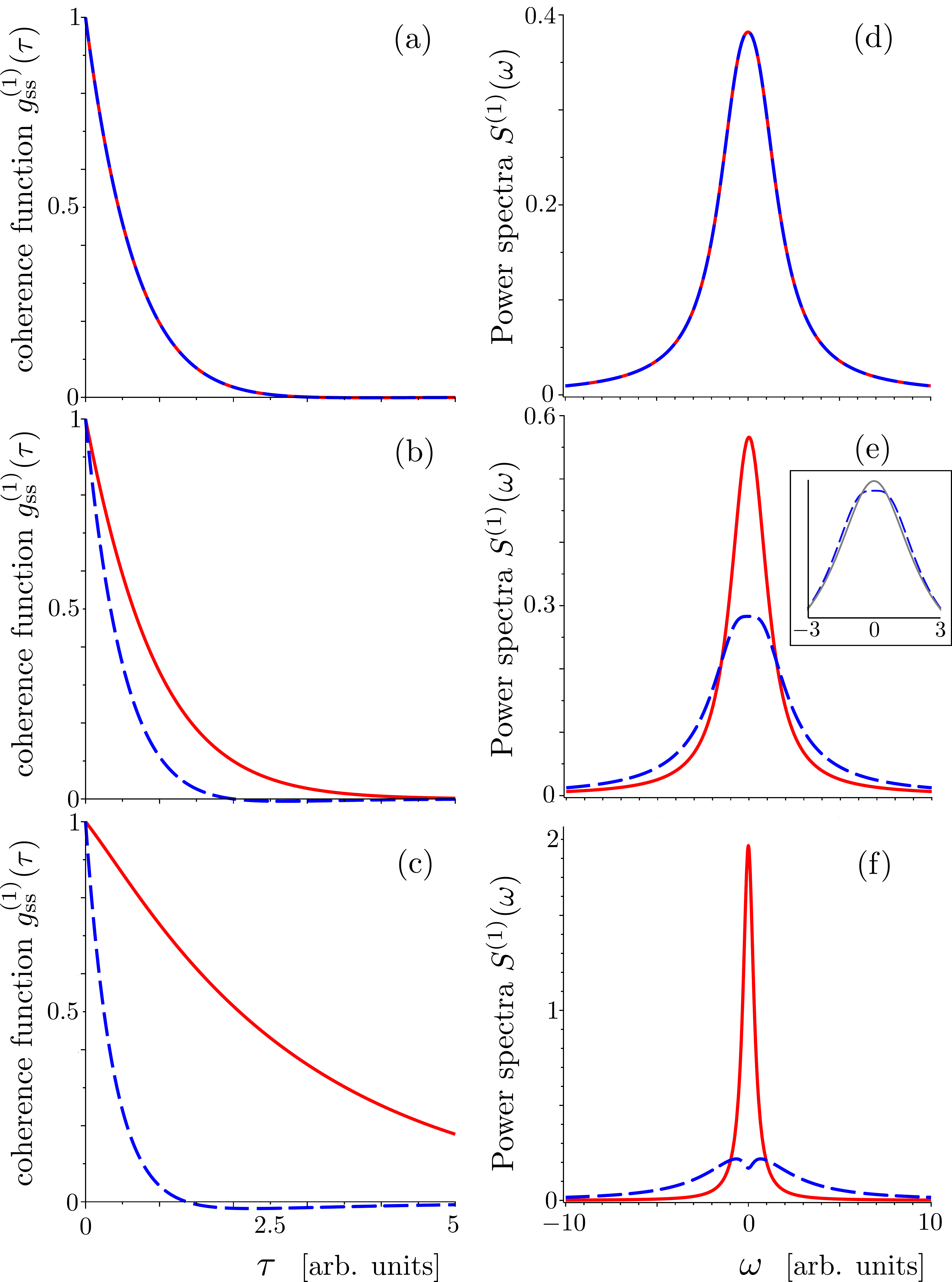}
\caption{First-order coherence function $g^{(1)}_{\rm ss}(\tau)$
[panels (a)-(c)] and power spectra $S^{(1)}(\omega)$ [panels (d)-(f)] of
the supermodes $\hat c_1$ (red solid curves) and  $\hat c_2$ (blue
dashed curves), according to Eqs.~(\ref{g1}) and (\ref{Sc1c2}),
respectively, of the bimodal cavity for various values of
incoherent mode coupling rate $\gamma_{12}$: (a) [(d)]
$\gamma_{12}=0$ [arb. units]; (b) [(e)]
$\gamma_{12}=\gamma_{12}^{\rm EP_1}=1$ [arb. units]; and (c) [(f)]
$\gamma_{12}=2$ [arb. units]. The reference frame is rotating at
the central cavity frequency $\bar\omega$. The remaining system
parameters are the same as in Fig.~\ref{fig1}. The spectra of the
supermodes $\hat c_{1}$ and $\hat c_{2}$ reveal the squared Lorentzian lineshape
at the second-order EP, characterized by a {\it plateau} at the top of the curve  [see panel (e) blue dashed curve].
The inset in panel (e) shows the best Lorentzian fitting, represented by a sum of two Lorentzians (gray solid curve) of the spectrum at the EP (blue dashed curve), which highlights the distinctiveness of the squared Lorentzian. Thus, one can experimentally witness the presence of the EP by studying the power spectra for the model under consideration.
Moreover, with
increasing values of  the incoherent mode-coupling rate
$\gamma_{12}>\gamma_{12}^{\rm EP_1}$, the supermode $\hat c_2$
acquires a volcanic cone shape ( blue dashed curve), indicating that its spectrum is
represented by the difference of  two  Lorentzians [panel (f)].
 }\label{fig2}
\end{figure}

\begin{figure}[h!] 
\includegraphics[width=0.465\textwidth]{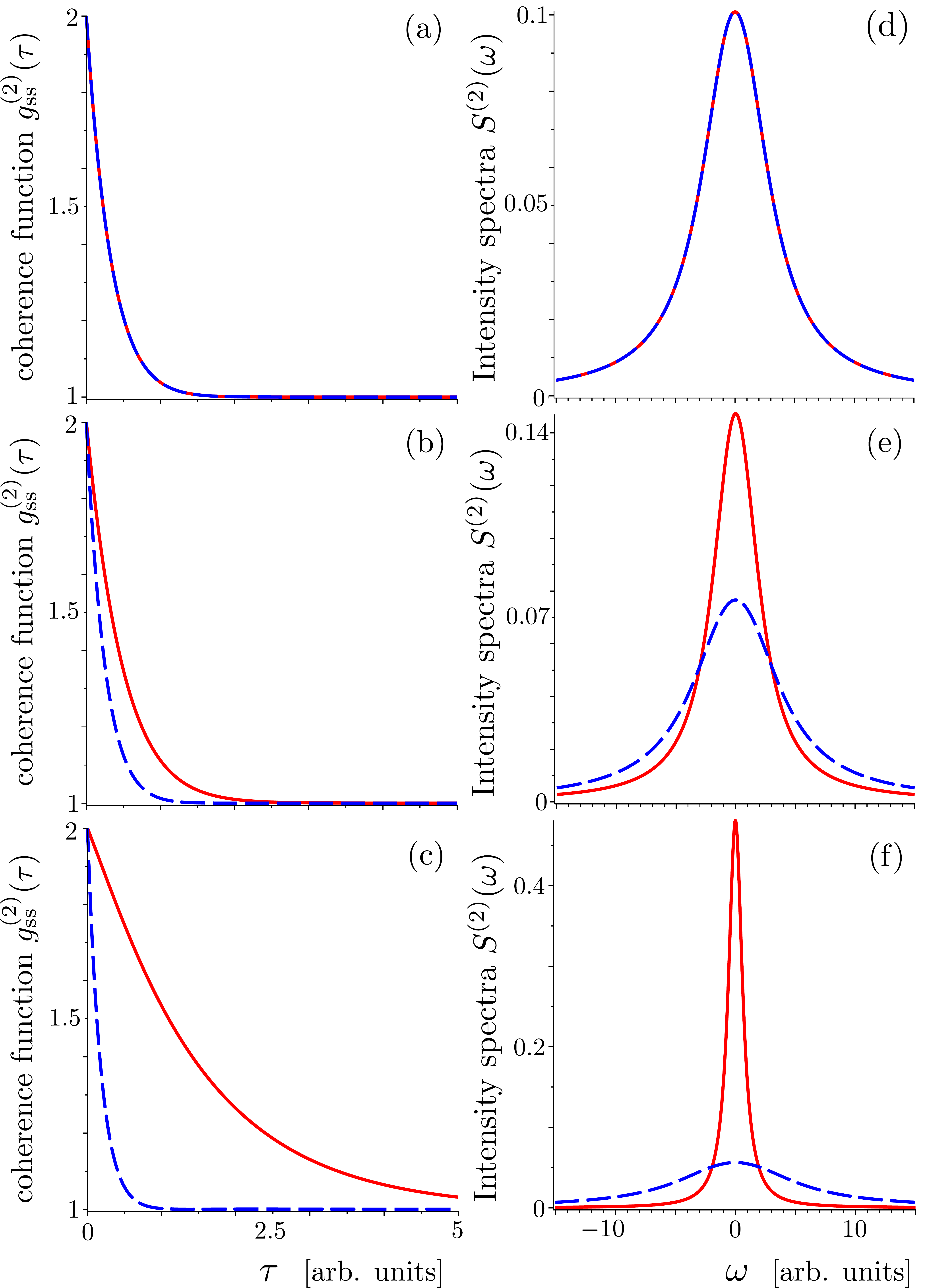}
\caption{Second-order coherence function $g^{(2)}_{\rm ss}(\tau)$
[panels (a)-(c)] and intensity-fluctuation spectra
$S^{(2)}(\omega)$ [panels (d)-(f)] of the supermodes $\hat c_1$
(red solid curves) and  $\hat c_2$ (blue dashed curves), according
to Eqs.~(\ref{g2g1}) and (\ref{s2}), respectively, of the bimodal
cavity for various values of incoherent mode coupling rate
$\gamma_{12}$. The system parameters for each panel are the same
as in Fig.~\ref{fig2}.
 }\label{fig3}
\end{figure}

The presence of  an EP of  second order in Eq.~(\ref{HEP})
can be signalled by a squared Lorentzian lineshape in the power
spectrum of the $\cal PT$-symmetric modes $\hat
c_{1,2}$~\cite{Yoo2011,Sweeney2019,Pick2017,Arkhipov2019b}.
Indeed, the anti-$\cal PT$-symmetric modes $\hat a_{1,2}$ generate two
independent spectra, which are not coupled. Thus, it is impossible
to see spectral lines merging which are highlighted by a squared
Lorentzian for these modes. However, the spectra of the
$\cal PT$-symmetric supermodes $\hat c_{1,2}$ can demonstrate
 the modes coalescence effect, {characterized by the appearance of a {\it plateau} at the top of the lineshape curve.}

The power spectrum $S^{(1)}(\omega)$ can be expressed via the
first-order coherence function $ g^{(1)}_{\rm ss}(\tau)$ as
\begin{equation} \label{Sw} 
S^{(1)}(\omega)=\frac{1}{\pi}{\rm Re}\int\limits_0^\infty
g^{(1)}_{\rm ss}(\tau)\exp(i\omega\tau){\rm d}\tau,
\end{equation}
Mathematically, $S^{(1)}(\omega)$ is, thus, the Fourier transform
of the coherence function $g^{(1)}_{\rm ss}(\tau)$, and, roughly
speaking, indicates the response of the system to the injection of
one particle at a frequency $\omega$.

The coherence function $g^{(1)}_{\rm ss}(\tau)$, for the supermodes
$\hat c_{1,2}$, with the help of Eq.~(\ref{RES1}), is found as:
\begin{equation}\label{g1} 
g^{(1)}_{\hat c_{1},\hat
c_2}(\tau)=\frac{\exp\left(-\frac{\gamma\tau}{2}-i\bar\omega\tau\right)}{D}\left(D\cosh\frac{D\tau}{2}\pm\gamma_{12}\sinh\frac{D\tau}{2}\right).
\end{equation}
The incoherent power spectra of the supermodes $\hat c_{1,2}$,
thus, take the form
\begin{equation}\label{Sc1c2} 
S^{(1)}_{\hat c_1,\hat c_2}(\omega)=\frac{1}{\pi
D}\left[\frac{K_{+}(D\mp\gamma_{12})}{\Omega^2+K_+^2}+\frac{K_{-}(D\pm\gamma_{12})}{\Omega^2+K_-^2}\right],
\end{equation}
where  $\Omega=\omega-\bar\omega$, and $K_{\pm}=(\gamma\pm D)/2$.

In Fig.~\ref{fig2}, we plot both the coherence function
$g^{(1)}_{\hat c_{1},\hat c_2}(\tau)$ and power spectra $S^{(1)}_{\hat
c_{1},\hat c_2}(\omega)$ for the supermodes $\hat c_1$ and $\hat
c_2$. Away from an EP, the coherence functions  (power spectra)
are a combination of two exponents (Lorentzians) for both fields
$\hat c_1$ and $\hat c_2$. When $\gamma_{12}=0$, the functions
$g^{(1)}_{\hat c_{1},\hat c_2}(\tau)$ and spectra $S^{(1)}_{\hat
c_{1},\hat c_2}(\omega)$  are identical, and the system is in the
exact (broken) \PT (anti-\PT) symmetric phase of the NHH. But if
the losses $\gamma$ are sufficiently large, the two cavity
resonances $\omega_1$ and $\omega_2$ might not be resolved [see
Fig.~\ref{fig2}(d)]. At the EP $\gamma_{12}=\gamma_{12}^{\rm EP}$,
 the coherence functions attain a nonexponential form:
\begin{equation}\label{g1ep} 
g^{(1)}_{\hat c_{1},\hat c_2}(\tau) =
\exp\left(-\frac{\gamma\tau}{2}-i\bar\omega\tau\right)\left(1\pm\frac{\gamma_{12}\tau}{2}\right).
\end{equation}
We note that  distinguishing the exponential from nonexponential
behavior could be hard in practice [see Fig.~\ref{fig2}(b)]. On
the other hand, at the EP, the two spectra become a combination of
the Lorentzian and squared Lorentzian lineshapes [see
Fig.~\ref{fig2}(e)]. Indeed, the power spectra $S^{(1)}_{\hat c_1,\hat
c_2}$ at the EP  become:
\begin{equation}\label{SEP} 
S_{\hat c_1,\hat c_2}^{\rm EP_1}(\omega)=\frac{1}{\pi}\frac{4}{\gamma^2+4\Omega^2}\left(\gamma\mp\gamma_{12}\pm\frac{2\gamma^2\gamma_{12}}{\gamma^2+4\Omega^2}\right),
\end{equation}
where $\Omega$ is given in Eq.~(\ref{Sc1c2}).

For larger values of $\gamma_{12}>\gamma_{12}^{\rm EP}$, the first
mode $\hat c_1$ experiences further amplification, whereas the
second mode $\hat c_2$ encounters an increased damping. Notice
also that the $\hat{c}_2$ mode, representing the difference of two
Lorentzians, is no longer characterized by a single maximum for
sufficiently large $\gamma_{12}$ [see Fig.~\ref{fig2}(c)]. The
Lorentzian subtraction can lead to a substantial decrease of the
spectral signal at the central cavity frequency $\bar\omega$,
meaning that the energy is completely transferred from the mode
$\hat c_2$ to $\hat c_1$. The latter leads to the observation of
 electromagnetically induced absorption in the
system~\cite{Zhang2020}. Clearly, the presence of the incoherent
mode coupling gives rise to a number of nontrivial phenomena in
the system, which are related to both anti-\PT and \PT symmetric
systems.
\begin{figure} 
\includegraphics[width=0.34\textwidth]{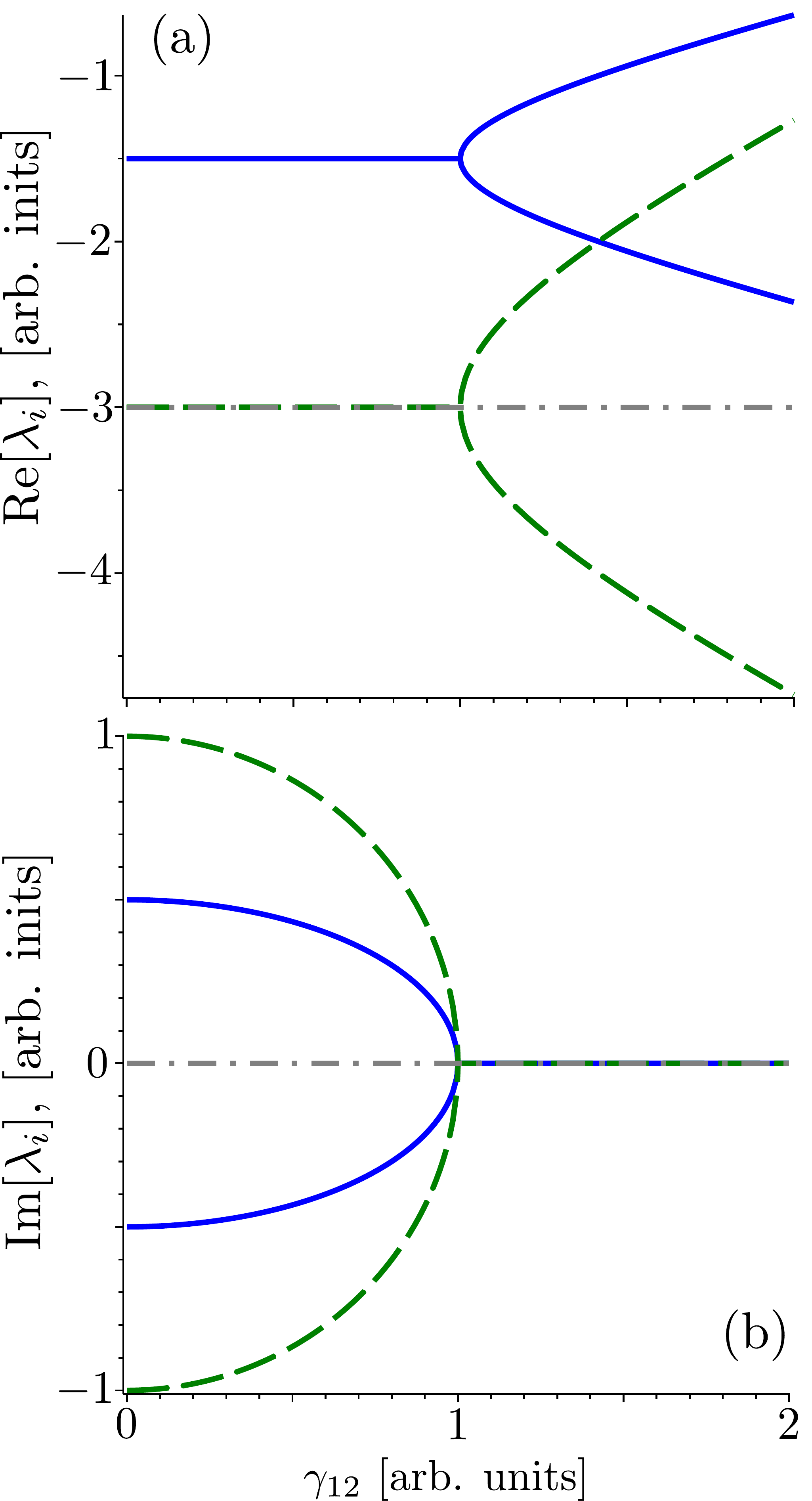}
\caption{(a) Real  and (b) imaginary parts of the eigenvalues
$\lambda$ of the Liouvillian, determined from the evolution matrix
$\boldsymbol{M}$: $\lambda_{1,2}$  (green dashed curves), and
$\lambda_{3,4}$ (grey dash-dotted lines), according to
Eq.~(\ref{Meig}), versus the incoherent coupling rate
$\gamma_{12}$. The chosen system parameters are the same as in
Fig.~\ref{fig1}. For comparison, the Liouvillian eigenvalues
defined by the NHH from Fig.~\ref{fig1} are shown as blue solid
curves. Thus, for a given subspace of the Liouvillian eigenspace,
determined by the NHH and evolution matrix $M$, the EP
$\gamma^{\rm EP}$, in Eq.~(\ref{HEP}), becomes a LEP of the second
and third orders, respectively. }\label{fig4}
\end{figure}

\subsubsection{Liouvillian exceptional point of  third order and cubic Lorentzian intensity-fluctuation spectra}
According to Eq.~(\ref{g2g1}), the second-order coherence function
$g^{(2)}(\tau)$ at the steady state is determined by the coherence
function $g^{(1)}(\tau)$. By exploiting Eqs.~({\ref{g2g1}}) and
(\ref{g1}), one can arrive to an explicit expression for the
function $g^{(2)}_{\hat c_{1},\hat c_2}(\tau)$. Namely, at the EP
in Eq.~(\ref{HEP}), from Eq.~(\ref{g1ep}), one obtains
\begin{equation}\label{g2ep} 
g^{(2)}_{\hat c_{1},\hat
c_2}(\tau)=1+\exp\left(-{\gamma\tau}\right)\left(1\pm\frac{\gamma_{12}\tau}{2}\right)^2,
\end{equation}
meaning that the function  $g^{(2)}_{\hat c_{1},\hat c_2}(\tau)$
signals the presence of a LEP of the third order. This third-order
LEP is also reflected by the cubic Lorentzian lineshape in the
intensity-fluctuation spectra, which is defined by the
second-order coherence function $g^{(2)}(\tau)$ at the steady
state as follows
\begin{equation}\label{s2} 
S^{(2)}(\omega)=\frac{1}{\pi}{\rm
Re}\int\limits_0^{\infty}\left[g^{(2)}(\tau)-1\right]\exp(i\omega\tau){\rm
d}\tau.
\end{equation}
By combining Eqs.~(\ref{g2ep}) and (\ref{s2}), one easily arrives
at the following expressions for the intensity-fluctuation spectra
(or noise spectra) for the supermodes at the EP:
\begin{equation}\label{s2ep} 
S^{(2)}_{\hat c_1,\hat
c_2}(\omega)\equiv\frac{\gamma\mp\gamma_{12}}{\omega^2+\gamma^2}-\frac{\gamma\gamma_{12}(3\gamma_{12}\mp4\gamma)}{2(\omega^2+\gamma^2)^2}+\frac{2\gamma^3\gamma_{12}^2}{(\omega^2+\gamma^2)^3}.
\end{equation}
As predicted by Eq.~(\ref{s2ep}), at the EP
$\gamma=\gamma_{12}^{\rm EP}$ of Eq.~(\ref{HEP}) the
intensity-fluctuation spectra are cubic Lorentzians [see
Figs.~\ref{fig3}(d)-\ref{fig3}(f)]. Experimentally, resolving the
exact cubic lineshape might be challenging due to the simultaneous
presence of the squared and cubic Lorentzians terms in
Eq.~(\ref{s2ep}). Thus, other techniques and methods might be
required to precisely detect it~\cite{Arkhipov2020}. As
Fig.~\ref{fig3} shows, the intensity-fluctuations decrease
(increase) for the mode $\hat c_1$ ($\hat c_2$), in accordance
with the power spectra in Fig.~\ref{fig2}.

\subsubsection{Liouvillian exceptional point of third order explicitly defined from the higher-order field moments matrix}
{From Eqs.~(\ref{g1}) and (\ref{g2g1}) one can even explicitly  identify
the Liouvillian eigenvalues, which determine the second-order
coherence function $g^{(2)}(\tau)$, and the merging of which gives rise to third-order LEP.} 
{We have already shown in Fig.~\ref{fig3} how a third-order
	LEP gives rise to the cubic Lorentzian in the
	intensity-fluctuation spectra, along with a quadratic-time
	dynamics for the second-order coherence function $g^{(2)}(\tau)$
	in Eq.~(\ref{g2ep}). The system is $U(1)$ symmetric, that is, the
	Liouvillian is invariant under any phase shift $\phi$ of the boson
	operators $\hat a_j\to\hat a_j\exp\left(i\phi\right)$, $j=1,2$.
	Thus, the time  dynamics of the  coherence function
	$g^{(2)}(\tau)$ is completely determined by the second-order
	moments of the fields of the form $\langle\hat a_k^{\dagger}\hat
	a_l\rangle$, $k,l=1,2$. The dynamics of such moments is determined
	by the corresponding Liouvillian eigenspace. Therefore, the
	knowledge of the time evolution of the second-order  moments
	$\langle\hat a_k^{\dagger}\hat a_l\rangle$ can reveal the presence
	of the third-order LEP. }
	
{ The dynamics of the second-order moments can be described by the averages of
$\boldsymbol{\hat a}=[(\hat
a_1^{\dagger},\hat a_2^{\dagger})\otimes(\hat a_1,\hat a_2)]^T$.
Indeed,}
\begin{equation}\label{h} 
\frac{\rm d}{{\rm d}t}\langle \boldsymbol{\hat
a}\rangle=\boldsymbol{M}\langle \boldsymbol{\hat a}\rangle+n_{\rm
th}\boldsymbol{b}.
\end{equation}
The evolution matrix $M$ for this vector of the averages  $\langle
\boldsymbol{\hat a}\rangle$ reads as
\begin{equation}\label{M} 
\boldsymbol{M}=\frac{1}{2}\begin{pmatrix}
-2\gamma & -\gamma_{12} & -\gamma_{12} & 0 \\
-\gamma_{12} & 2i\Delta-2\gamma & 0 & -\gamma_{12} \\
-\gamma_{12} & 0 & -2i\Delta-2\gamma & -\gamma_{12} \\
0 & -\gamma_{12} & -\gamma_{12} & -2\gamma
\end{pmatrix},
\end{equation}
and the thermal noise vector is
$\boldsymbol{b}=[\gamma,\gamma_{12},\gamma_{12},\gamma]^T$.

In the supermode basis, the evolution matrix $\boldsymbol{N}$ and
the noise vector $\boldsymbol{d}$ for the vector $\boldsymbol{\hat
c}=[(\hat c_1^{\dagger},\hat c_2^{\dagger})\otimes(\hat c_1,\hat
c_2)]^T$ are easily found via the transformation
\begin{equation}\label{Nd} 
\boldsymbol{N}=\boldsymbol{T}\boldsymbol{M}\boldsymbol{T}^{-1},
\quad \boldsymbol{d}=\boldsymbol{T}\boldsymbol{b},
\end{equation}
where the $4\times4$ transformation matrix $\boldsymbol{T}$ is
given by
\begin{equation} 
\boldsymbol{T}=\frac{1}{2}\begin{pmatrix}
1 & -1 \\
1 & 1
\end{pmatrix}\otimes
\begin{pmatrix}
1 & -1 \\
1 & 1
\end{pmatrix}.
\end{equation}
From Eqs.~(\ref{h}) and (\ref{Nd}) it is evident that the dynamics
for the vector of the averages $\langle \boldsymbol{\hat
a}\rangle$ ($\langle\boldsymbol{\hat c}\rangle$) cannot, in
general,  possess anti-\PT (${\cal PT}$)-symmetry because of the
presence of the thermal noise in the form of the vector
$\boldsymbol{b}$ ($\boldsymbol{d}$). Nevertheless, the matrix
$i\boldsymbol{M}$ ($i\boldsymbol{N}) $ is anti-\PT (${\cal
PT}$)-symmetric. Indeed, the parity operator $\cal P$ for the
vector of the operators $\boldsymbol{\hat a}, \boldsymbol{\hat c}$
becomes
\begin{equation}\label{P} 
{\cal P}=\begin{pmatrix}
0 & 1 \\
1 & 0
\end{pmatrix}\otimes
\begin{pmatrix}
0 & 1 \\
1 & 0
\end{pmatrix}.
\end{equation}
With the help of Eq.~(\ref{P}) one can easily check that
\begin{eqnarray}\label{MN} 
&{\cal PT}\left(i\boldsymbol{M}\right)\left({\cal PT}\right)^{-1}=-i\boldsymbol{M},&\nonumber \\
& {\cal PT}\left(i\boldsymbol{N'}\right)\left({\cal
PT}\right)^{-1}=i\boldsymbol{N'},&
\end{eqnarray}
where the modified matrix $\boldsymbol{N'}$ is obtained from
$\boldsymbol{N}$ by applying the gauge transformation in
Eq.~(\ref{gauge}). The inclusion of the imaginary prefactor $i$ in
the matrices $\boldsymbol{M}$  and $\boldsymbol{N'}$, in
Eq.~(\ref{MN}), ensures that the l.h.s. of the equations of motion
for the vectors of the operators $\langle\boldsymbol{\hat
a}\rangle$ and $\langle\boldsymbol{\hat c}\rangle$ remain
unchanged under \PT transformation. Thus, in the absence of
thermal photons $n_{\rm th}=0$, the dynamics for the averaged
vector of operators $\langle \boldsymbol{\hat a}\rangle$ ($\langle
\boldsymbol{\hat c}\rangle$) restores the  same anti-\PT (${\cal
PT}$)-symmetry imposed by the effective NHH on quantum fields in
Eq.~(\ref{HeffMa}) [(\ref{Hc1c2ef})]. As a result, the LEP of
third order, determined from the evolution matrix $\boldsymbol{M}$
($\boldsymbol{N}$), {\it becomes directly associated with anti-\PT
(${\cal PT}$)-symmetry breaking.}

The eigenvalues of the matrix $\boldsymbol{M}$, and thus of the
matrix $\boldsymbol{N}$, are found as follows
\begin{equation}\label{Meig} 
\lambda_{1,2}=-2i\nu_{12}, \quad \lambda_{3,4}=-\gamma,
\end{equation}
where $\nu_{1,2}$ are the eigenvalues of the NHH given in
Eq.~(\ref{nu12}). We plot these eigenvalues in Fig.~\ref{fig4}. At
the EP $\gamma_{12}^{\rm EP}$, the algebraic multiplicity of the
eigenvalue $\lambda=-\gamma$ equals four, whereas geometric
multiplicity is three. In other words, there is a coalescence of
three Liouvillian eigenvectors, which are determined by the
moments of the operators in the vectors $\boldsymbol{\hat a}$ and
$\boldsymbol{\hat c}$, but the derivation of their explicit form
might require other approaches~\cite{Honda2010,Teuber2020}. We
also note that this finding of the LEP of third order for a
second-order HEP in the space of the vector $\boldsymbol{\hat c}$
has already been observed in the single-photon regime for a
similar system~\cite{Arkhipov2020}.

{The presence of higher-order EPs in a system is
usually associated with the enhanced system sensitivity to
external perturbations in the vicinity of the EPs~\cite{Ozdemir2019,Wiersig2014}. This
system's enhanced spectral response $\Delta\omega$ near an EP of
an $n$th-order to a perturbation $\epsilon$ scales as
$\Delta\omega\sim\epsilon^{\frac{1}{n}}$. Remarkably, the
system spectral sensitivity around the LEP of the third-order can
remain the same as it is near the second-order LEP. That is, the
Liouvillian eigenvalues split near the third-order LEP, as
$\Delta\lambda\sim\sqrt{\epsilon}$, not as cubic root as one might
expect. This squared-root dependence on perturbation around the
third-order LEP  arises from the system symmetry and the nature of
the applied perturbation. }

{As it was explained earlier, the
eigenvalues $\lambda_{1,2,3}$ Eq.~(\ref{Meig}) belong to the
corresponding $U(1)$ Liouvillian eigenspace. As such, any
perturbation of a single system parameter preserves the $U(1)$ symmetry of the system,
e.g., $\gamma\to\gamma+\epsilon$. Consequently, the eigenvalue $\lambda_3$ (and the corresponding Liouvillian eigenmatrix) remain real (and Hermitian) under such a perturbation. Since $\lambda_3$ never acquires a imaginary part for such perturbations, only the complex eigenvalues $\lambda_{1,2}$, along with their eigenstates, induce a line
splitting in the intensity-fluctuation spectrum around the
third-order LEP. Thus, the spectral
response to such external perturbations scales only as the squared
root at the third-order LEP. }

{This result can be generalized to higher-order LEPs.
Since any Liouvillian eigenvalue come in conjugate pairs, and given the $\mathcal{PT}$-symmetric structure of the system, given the coalescence of
$(2k+1)$ eigenvalues, one eigenvalue must always remain purely real in the vicinity of the LEP. 
Such an eigenvalue cannot contribute to the enhanced system spectral response under a perturbation. 
We conclude that, for the studied system, for any
LEPs of odd order $(2k+1)$, the system enhanced sensitivity
scales at most as $\epsilon^{\frac{1}{2k}}$ around the LEP. }

Finally, we would like to stress the mentioned difference in the
dynamics of the higher-order field moments imposed by the
Liouvillian and NHH. To show this explicitly we write the
corresponding evolution matrix $\boldsymbol{M_{\rm NHH}}$ for the
vector of the operators $\boldsymbol{\hat a}$, derived from the
NHH in Eq.~(\ref{HeffMa}) as follows
\begin{equation}\label{MNHH} 
\boldsymbol{M_{\rm NHH}}=\frac{1}{2}\begin{pmatrix}
0 & \gamma_{12} & -\gamma_{12} & 0 \\
\gamma_{12} & 2i\Delta & 0 & -\gamma_{12} \\
-\gamma_{12} & 0 & -2i\Delta & \gamma_{12} \\
0 & -\gamma_{12} & \gamma_{12} & 0
\end{pmatrix}.
\end{equation}
A comparison of Eqs.~(\ref{M}) and (\ref{MNHH}) demonstrates that,
indeed, the evolution imposed on the same operators is different
in the Liouvillian and NHH formalisms. The eigenvalues of the
matrix $\boldsymbol{M_{\rm NHH}}$ are similar to those in
Eq.~(\ref{Meig}) but shifted by the value of $\gamma$, i.e.,
$\lambda_{\rm NHH}=\lambda_{\cal L}+\gamma$, where $\lambda_{\cal
L}$ are given in Eq.~(\ref{Meig}). Additionally, the inhomogeneous
term $\boldsymbol{b}$, arising from the thermal noise, is absent
in the equations of motion for the operators $\boldsymbol{\hat a}$
in the NHH formalism. Therefore, although the HEPs and LEPs can
coincide for the same field moments, the system eigenspectra and
dynamics are different in both
formalisms~\cite{Minganti2019,Arkhipov2020}.

\section{Conclusions}\label{V}

In this work, we have demonstrated that for a dissipative linear
bosonic system, whose effective NHH exhibits a  HEP of any finite
order $n$, determined by the first-order field moments, its
corresponding LEPs can become at least of order $m\geq 2k(n-1)+1$,
for $k=\frac{1}{2},1,2,3,\dots$, which can be determined by
higher-order field moments, accordingly. These higher-order field
moments are directly related to the {\it normally-ordered} higher-order coherence
functions via the quantum regression theorem.

Thus, we have shown how the coherence functions can offer a
convenient tool to probe extreme system sensitivity to external
perturbations in the vicinity of higher-order LEPs.

As an example, we have studied a bosonic system of a bimodal
cavity with incoherent mode coupling to reveal its higher-order
LEPs. In particular, we analyze second and third-order LEPs of
such systems arising from a HEP of  second order by calculating
the first- and second-order coherence functions, respectively.
Also, we calculate the corresponding power and
intensity-fluctuation spectra to reveal their squared and cubic
Lorentzian expressions, accordingly. 
{Moreover, our analysis of such two-mode systems indicates that, for the LEPs of an odd order $(2k+1)$,
the system enhanced sensitivity to external perturbations $\epsilon$ scales at most as $\epsilon^{\frac{1}{2k}}$.}

We also reveal the anti-$\cal PT$- and $\cal PT$-symmetries of the NHH, which
connect the presence of LEPs with spontaneous breaking  of such
symmetries in a bimodal cavity with incoherent mode coupling.
Moreover, we showed the possibility of switching between the \PT
and anti-\PT symmetries of the studied bosonic linear systems by
applying an additional tunable linear coupler to two supermodes of the
system output. { By means of such coupler one can, thus,  transform the initial loss-loss
dynamics for the supermodes to the equivalent system with the balanced
gain-loss evolution.}
In case of optical systems, this transformation can
be implemented with a single tunable beam splitter.

We note that usually EPs have been studied in two- or multiparty
systems with gain and loss (i.e., lossy driven systems). Here, we
have analyzed EPs in a multiparty system without gain, but instead
with its subsystems exhibiting losses with different rates. Such a
system effectively leads to a model of a lossy-driven system.

In conclusion, we believe that our work has shown that the concept
of quantum EPs, as defined via degeneracies of Liouvillians, is
not only of a pure theoretical interest. We have demonstrated
explicitly that the formalism of Ref.~\cite{Minganti2019} can be
tested experimentally at least for general quantum linear bosonic
systems. In particular, LEPs can indeed be identified by measuring
coherence functions or the power and intensity fluctuation
spectra.

\section*{Acknowledgments}

I.A. thanks the Grant Agency of the Czech Republic (Project
No.~18-08874S), and Project no.
CZ.02.1.01\/0.0\/0.0\/16\_019\/0000754 of the Ministry of
Education, Youth and Sports of the Czech Republic. A.M. is
supported by the Polish National Science Centre (NCN) under the
Maestro Grant No. DEC-2019/34/A/ST2/00081. F.M. is supported by
the FY2018 JSPS Postdoctoral Fellowship for Research in Japan.
F.N. is supported in part by: NTT Research, Army Research Office
(ARO) (Grant No. W911NF-18-1-0358), Japan Science and Technology
Agency (JST) (via the CREST Grant No. JPMJCR1676), Japan Society
for the Promotion of Science (JSPS) (via the KAKENHI Grant No.
JP20H00134, and the JSPS-RFBR Grant No. JPJSBP120194828),
and the Grant No. FQXi-IAF19-06 from the Foundational Questions
Institute Fund (FQXi), a donor advised fund of the Silicon Valley
Community Foundation.

\appendix
\section{Liouvillian $\cal L$ for supermodes}\label{AA}
Here we show an explicit form of the Liouvillian $\cal L$ for the supermodes
$\hat c_{1,2}$ in Eq.~(\ref{c1c2}).

Considering a general Liouvillian, given in Eq.~(\ref{L}), for two
modes with arbitrary damping matrix $\gamma_{jk}$ and free
Hermitian Hamiltonian in Eq.~(\ref{Hfree}),  after applying the
transformations in Eq.~(\ref{c1c2}), one arrives at the
transformed master equation given by
\begin{eqnarray}\label{Ltheta}
\frac{{\rm d}}{{\rm d}t}\hat\rho={\cal L}\hat\rho &&= -i[\hat H_{c_1}+\hat H_{c_2}+\hat H_{c_1,c_2},\hat\rho]\hat\rho \nonumber \\
&&+{\cal L}_{c_1}\hat\rho+{\cal L}_{c_2}+{\cal
L}_{c_1,c_2}\hat\rho+{\cal L}_{c_2,c_1}\hat\rho,
\end{eqnarray}
where the coherent part includes:
\begin{eqnarray}
\hat H_{c_1}&=&\left(\omega_1\cos^2\theta+\omega_2\sin^2\theta\right)\hat c_1^{\dagger}\hat c_1, \nonumber \\
\hat H_{c_2}&=&\left(\omega_1\sin^2\theta+\omega_2\cos^2\theta\right)\hat c_2^{\dagger}\hat c_2, \nonumber \\
\hat H_{c_1,c_2}&=&\frac{1}{2}(\omega_1-\omega_2)\sin2\theta(\hat
c_1^{\dagger}\hat c_2+\hat c_2^{\dagger}\hat c_1),
\end{eqnarray}
and the incoherent Lindbladian part is:
\begin{equation}
{\cal L}_{c_m}\hat\rho=\frac{n_{\rm th}+1}{2}A_m{\cal D}\left[\hat
c_{m}\right]+\frac{n_{\rm th}}{2}A_m{\cal D}\left[\hat
c_{m}^{\dagger}\right],
\end{equation}
for $m=1,2$, and
\begin{equation}
{\cal L}_{c_j,c_k}\hat\rho=\frac{n_{\rm th}+1}{2}A_{jk}{\cal
D}\left[\hat c_{j}\hat c_k^{\dagger}\right]+\frac{n_{\rm
th}}{2}A_{jk}{\cal D}\left[\hat c_{j}^{\dagger}\hat c_k\right],
\end{equation}
with $j,k=1,2$ and $j\neq k$. Moreover we have denoted
\begin{eqnarray}
A_1&=&\gamma_{11}\cos^2\theta+\gamma_{22}\sin^2\theta-\bar\gamma_{12}\sin2\theta, \nonumber \\
A_2&=&\gamma_{11}\sin^2\theta+\gamma_{22}\cos^2\theta+\bar\gamma_{12}\sin2\theta, \nonumber \\
A_{12}&=&\gamma_-\sin2\theta+\gamma_{12}\cos^2\theta-\gamma_{21}\sin^2\theta, \nonumber \\
A_{21}&=&\gamma_-\sin2\theta+\gamma_{21}\cos^2\theta-\gamma_{12}\sin^2\theta.
\end{eqnarray}
where $\bar\gamma_{12}=(\gamma_{12}+\gamma_{21})/2$, and
$\gamma_-=(\gamma_{11}-\gamma_{22})/2$. Now, for the considered
symmetric damping matrix ($\gamma_{11}=\gamma_{22}$ and
$\gamma_{12}=\gamma_{21}$), by taking $\theta=\pi/4$, the Lindblad
operators in Eq.~(\ref{Ltheta}) become diagonalized. Thus, the
initially lossy system with incoherent mode coupling becomes a
lossy system with coherent mode coupling with an NHH, given in
Eq.~(\ref{Hc1c2ef}).


%

\end{document}